\documentstyle[12pt,aasms4]{article}
  
\newcommand{\nix}{$\cdot\cdot\cdot$}
   
\begin{document}
    
\title{Are the Nuclei of Seyfert 2 Galaxies Viewed Face-On?}

\author{K. A. Weaver}
\affil{Johns Hopkins University, Department of Physics and Astronomy,
Homewood Campus, 3400 North Charles Street, Baltimore,
MD 21218-2695; kweaver@pha.jhu.edu}
\and
\author{C. S. Reynolds}
\affil{JILA, Campus Box 440, University of Colorado,
Boulder, CO 80309-0440}

\begin{abstract}
  
We show from modeling the Fe K$\alpha$ line in the $ASCA$ spectra of four
X-ray bright narrow emission line galaxies (Seyfert types 1.9 and 2) that
two equally viable physical models can describe the observed line
profile.  The first is discussed by Turner et al.\ (1998) and consists of
emission from a nearly pole-on accretion disk.  The
second, which is statistically preferred,  
is a superposition of emission from an accretion disk viewed at an
intermediate inclination of $\sim$48$^{\circ}$ and a distinct,
unresolved feature that presumably originates some distance from the
galaxy nucleus.  The intermediate inclination is entirely consistent with
unified schemes and our findings challenge recent assertions that Seyfert
2 galaxies are preferentially viewed with their inner regions face-on.
We derive mean equivalent widths for the narrow and disk lines of $<{\rm
EQW}_{\rm N}> = 60$ eV and $<{\rm EQW}_{\rm D}> = 213$ eV,
respectively.  The X-ray data are well described by a geometry in which our
view of the active nucleus intersects and is blocked by the outer edges
of the obscuring torus, and therefore do not require severe misalignments
between the accretion disk and the torus.

\end{abstract}
 
  
\section{Introduction}

For over a decade, the paradigm for active galactic nuclei (AGN) has rested
soundly on the unified model hypothesis, which posits that the only
difference between broad-line objects (e.g., Type 1 Seyfert galaxies) and
narrow-line objects (e.g., Type 2 Seyferts) is that in the former case our
line of sight evades obscuration by the optically thick torus surrounding
the nucleus, while in the latter, our line of sight intersects and is
blocked by the torus.  Assuming that the torus and inner accretion disk are
co-planar, this hypothesis makes testable predictions about the shape of
the Fe K$\alpha$ emission line profile, which is commonly very strong in
Seyfert galaxies and can yield important information about the orientation
of the system relative to our line of sight, as well as the geometry and
dynamics of the emitting matter.  In the standard picture of Seyfert
galaxies, when looking along or near the axis for a face-on system, we
expect the K$\alpha$ photons to arise mostly in an accretion disk resulting
in a line profile that is broadened by gravitational and Doppler effects
(e.g., George \& Fabian 1991; Laor 1991).  When looking in the equatorial
plane, the prediction of the standard model for the Fe K$\alpha$ line
depends on the optical depth and geometry of the obscuring matter.  If the
column density is small enough to let X-rays of energy greater than a few
keV go through, we would see a very broad, but relatively weak contribution
from the disk, and a relatively strong, narrow component from the obscuring
torus (Ghisellini, Haardt \& Matt 1994; hereafter GHM94, 
Krolik, Madau \& Zycki 1994; hereafter KMZ94).  In
all cases, a weak component from fluorescence in the clouds making up the
broad-line region may also be present (Leahy \& Creighton 1993).

The standard unified model has recently been brought into question by
studies of the detailed shape of the Fe K$\alpha$ emission line (Turner et
al.\ 1998, hereafter T98).  These authors argue that in Type 2 Seyfert
galaxies with small enough absorbing columns that the 6 keV nuclear
continuum can be seen directly, there are K$\alpha$ lines with enough
breadth to suggest that they are made in an accretion disk, but so narrow
and asymmetric that the inferred viewing direction is nearly polar and
consistent with the mean inclination angle for Seyfert 1s (Nandra et al.\ 
1997).  If these studies are borne out, it would mean that in some Type 2
Seyferts a relatively small amount of absorption is present, but lies on
the disk axis rather than solely in its equatorial plane, or that the disk
axis is severely misaligned with respect to the the galaxy axis and the
absorption arises in the galaxy instead of the torus.  Furthermore, since
Nandra et al. (1997) showed that Seyfert 1 nuclei also have iron line
profiles indicative of face-on disks, the T98 hypothesis would force us to
consider an {\it entire missing population} of AGN containing edge-on
disks.
 
This result has far-reaching consequences for studies of AGN and may force
us to re-examine, if not altogether abandon, many of our assumptions for
current unified models.  However, X-ray spectral data are often subject to
multiple interpretations.  The conclusion reached by T98 that Type 2 Seyfert 
galaxies are preferentially viewed with their central regions face-on is
based on data from the $ASCA$ satellite (Tanaka, Inoue \& Holt 1994);
primarily from spectral modeling of four X-ray selected Type 2 and 1.9
Seyfert galaxies (historically referred to as narrow emission line
galaxies or NELGs).  In this paper, we seek to determine how strongly
these data {\it require} the central regions of Seyfert 2s to be viewed
preferentially face-on by examining a physically plausible modeling of the
Fe K$\alpha$ line that is consistent with unified schemes.  We re-examine
the $ASCA$ data and use
an alternate modeling technique that allows significant contributions to
the Fe K$\alpha$ line from both the accretion disk and the torus.

\section{Data analysis and Results}

The $ASCA$ data are taken from the public archive.  
We try as much as possible to reproduce the T98 dataset
by selecting the same observations and adopting the same criteria for 
data selection.  The AGN in this sample
possess flux in excess of the
extrapolation of a uniformly absorbed power-law to energies below 2
keV (i.e., a ``soft excess'', T98).  To avoid complexities of modeling the
soft excess, we constrain our fits to data between 2 and 10 keV 
and use a baseline continuum model that consists of 
a power law (PL) attenuated by uniform, neutral absorption.  We list
the value of $\chi^2$ and the number of degrees of freedom 
for the PL fits in Table 1, column 3.  We do not quote 
model parameters for these baseline fits because they are similar to 
those quoted by Turner et al.\ (1997) and are unimportant for our study.

The Fe K$\alpha$ lines in these galaxies are strong and resolved 
with $ASCA$, and have been
examined in detail by other authors (T98, MCG$-$5-23-16: Weaver et al.\ 1997,
NGC 7314: Yaqoob et al.\ 1996, NGC 2110: Hayashi et al.\ 1996).  In Table 1
we show results for fits that include a single, unresolved (narrow)
Gaussian at a rest energy of 6.4
keV to represent fluorescence of neutral iron.  Including a Gaussian
improves all fits at $>>99\%$ confidence compared to the PL fits
(col.\ 3), thus confirming the significance of the Fe K$\alpha$ lines. 
Figure 1 shows the co-added line profile measured with the 
Solid-State Imaging Spectrometer (SIS) detectors (see 
also T98, Fig.\ 2).  The profile peaks strongly at the center,
which coincides with 6.4 keV (rest energy), and emission 
on either side of the core from $\sim$4.5 to 8 keV is clearly present.  
The line shape is, in fact, very   
different from some Seyfert 1 galaxies,
which show a strong red wing but little emission blueward of 
the line core.  One such example is the well-studied galaxy 
MCG$-$6-30-15 (Iwasawa et al.\ 1996).

We next test the standard accretion disk-line models used 
by T98 (Fabian et al.\ 1989).  These models assume
a Schwarzschild geometry and represent the disk emissivity as a power-law
function of disk radius, in this case $r^{-2.5}$.  We also fix the outer
disk radius, $r_{o}$, at 1000$r_{g}$ and fix the inner disk 
radius, $r_{i}$, at 6$r_{g}$, where $r_{g} = GM/c^2$. 
The free parameters are the
inclination of the disk normal to our line of sight ($\theta$) and the line
normalization.  For disk-line models, the shape of the line 
changes dramatically with disk inclination as shown in Figure 2. 
Results of the fits with the PL + disk-line model are 
presented in Table 2.
The standard Schwarzschild disk-line model yields a better fit compared 
to a narrow Gaussian for three of the four galaxies.  We derive 
a mean inclination for the accretion disk of $<\theta> = 38^{\circ}$ 
and a mean equivalent width for
the disk line of $<{\rm EQW}_{\rm D}> = 300$ eV.
Our inclination is somewhat larger than T98, who find $<\theta> =
20^{\circ}$ for these data. 

We have carefully examined possible reasons for the 
discrepancy between our derived inclinations and those of T98. 
These authors do not make clear whether they include a weak 
narrow component in the disk-line fits they quote, so we also 
performed the above fits including a weak, narrow Gaussian 
(${\rm EQW}_{\rm N} < 50$ eV).  For this case,
the only way to approximate low 
inclinations is by blueshifting the narrow component so that 
it is no longer located at the systemic velocity of the galaxy.
We can therefore achieve consistency with T98, but only with a model 
that differs slightly from the one these authors quote.  We
believe that the difference between our values of $\theta$ 
and those of T98 are not due to large, intrinsic uncertainties 
with a particular model (such that both
results actually overlap in some overlooked, statistical sense).  
Indeed, we argue that the T98 model is incomplete and, based 
on our careful examinations of the data, possibly non-physical.
 
We believe that a more physical composite model 
for the Fe K$\alpha$ line consists of 
a disk line and a narrow Gaussian with an 
unrestricted equivalent width (${\rm EQW}_{\rm N}$). 
The upper limit on ${\rm EQW}_{\rm N}$
of 50 eV that is assumed by T98 is a reasonable 
assumption for approximately face-on views if the narrow component arises
purely in a spherical distribution of clouds (such as the BLR) as predicted
from Monte Carlo simulations (Leahy \& Creighton 1993); however, the narrow
component may be stronger if the torus contributes, especially as the view
of the central region becomes more edge-on 
(KMZ94).  We therefore choose not to
restrict ${\rm EQW}_{\rm N}$ in our fits.  
Results for our composite model are 
listed in Table 3.  In all cases, the fits are
better than fits with a single disk-Line (Table 2), with probabilities 
of P = 0.01 to 0.001 of exceeding the F-statistic by chance. 
We derive $<\theta> = 48^{\circ}$, $<{\rm EQW}_{\rm N}> = 60$ eV,
and $<{\rm EQW}_{\rm D}> = 213$ eV.

In Table 4, we compare the accretion-disk parameters inferred for 
our disk-plus-torus model to those derived by T98 for a pure
disk-line description of the Fe K$\alpha$ line.
We derive systematically larger inclinations for the accretion 
disk and up to 50\% smaller values for $<{\rm EQW}_{\rm D}>$.
Both models are plotted together at full resolution in Figure 3.
For emission from a disk viewed at an intermediate inclination, 
significant flux is predicted blueward of the 
line core.  This signature is clearly present in the data 
(Figure 1) but is absent for the T98 ``pole-on'' 
disk model.  The two models are, however,  
qualitatively similar below   
6.4 keV.  Because of this similarity, X-ray data with 
the restricted bandpass and moderate energy resolution
of $ASCA$ (0.6 to 10 keV and $\ge130$ eV FWHM\footnote{The 
energy resolution of the SIS detectors has been degrading
steadily with time.}, 
respectively) make it difficult, if not impossible to
distinguish between the two cases.   The $ASCA$
data do not necessarily constrain the shape of the 
continuum {\it above} the line and so detecting
a blue wing can be highly sensitive to how 
the continuum is modeled.

\section{Discussion}

We have shown that the Fe K$\alpha$ emission lines in four X-ray selected 
Type 1.9 and 2 Seyfert galaxies are well modeled as the superposition
of two features, a narrow Gaussian at a rest energy of 6.4 keV and an
emission line from an accretion disk viewed at an intermediate inclination
angle of $\sim$48$^{\circ}$.  Statistically, this model describes the 
observed profiles better than a single, highly-peaked emission line from an
accretion disk that is seen approximately face-on (suggested by T98). 
We argue that current moderate-resolution X-ray data for
Seyfert galaxies cannot easily deconvolve the multiple
Fe K$\alpha$ emission components and so cannot necessarily distinguish 
between these scenarios without variability information or
information from other wavebands that suggests the
true origin of the line(s).  It must be noted that
Yaqoob et al.\ (1996) have found the iron line profile in NGC 7314 to be
variable in exactly the manner expected if there is a narrow component of
the line that originates at a large distance from the black hole.  However,
the comparatively low signal to noise of the time-resolved spectra
presented by Yaqoob et al.\ (1996) prevents a detailed decomposition of the
narrow and broad line components.

Our results suggest that an intermediate-inclination broad 
line contaminated by a
narrow component can mimic the predicted shape for emission from an
approximately face-on accretion disk. 
T98 dismissed the two-component model 
because they found that weak narrow components (EQW from $10-50$ eV) 
had no effect on their fits.  We argue that there is no
reason to restrict the strength of a contaminating iron line 
from the molecular torus to have an equivalent width less than 50 
eV.  First of all, we do not know the inclination 
of the system {\it a priori}.  There is in fact independent
evidence for an intermediate to edge-on orientation for at least one of 
the NELGs in our sample (MCG$-$5-23-16; Weaver, Krolik \& Pier 1998).
In the case of NELGs viewed 
at intermediate angles, we would expect our line of sight to  
intersect the edges of the torus just outside the 
half opening angle, which is consistent with the moderate 
line-of-sight column densities of $\sim 10^{22}$ to  
$\sim 10^{23}$ cm$^{-2}$.  If these measurements  
represented the mean column density through 
the torus, $<N_{\rm H}({\rm T})>$, then ${\rm EQW}_{\rm N}$
would be a few tens of eV for all inclination angles 
(GHM94) and would contribute
only a small fraction of the total line photons.   However,
the column density through the torus for other lines of
sight could easily be as high as 10$^{24-25}$ cm$^{-2}$. 
This would result in a larger $<N_{\rm H}({\rm T})>$ and   
EQWs of up to $\sim$100 eV for unblocked views of the 
continuum source, with the actual value depending
on the geometry of the torus (KMZ94; GHM94). 
We find $<{\rm EQW}_{\rm N}> = 60$ eV, which is 
well within this limit. 

For the disk-line component of the composite model, we derive
${\rm EQW}_{\rm D}$'s of $\sim$120 eV to $\sim$250 eV.  These 
are as much as 50\% 
smaller than those inferred for the pure disk-line 
model (this analysis and T98). 
If, as we suggest, multiple components {\it do} comprise the line, 
then this partly alleviates a fundamental 
problem that has been frequently 
pointed out in the literature: the observed EQWs of Fe K$\alpha$
lines in Seyfert galaxies are generally larger than the maximum 
predicted for an accretion disk (e.g., Reynolds \& Fabian 1997). 
A neutral disk inclined by 
$\theta = 50^{\circ}$ should produce an Fe K$\alpha$ 
line with an EQW of $\sim$120 to 170 eV, depending on the value 
assumed for the solar abundance of iron 
(George \& Fabian 1991; Reynolds, Fabian \& 
Inoue 1995).  For our model, three of four disk lines have EQWs 
larger than expected, but only by $\sim$50\%.  T98 find
all four EQWs to be larger than expected, by $\sim$50\%
to as much as $\sim$140\%.  Equivalent widths of more than
$\sim$200 eV can result if the disk is ionized
(Matt, Fabian, \& Ross 1993), although an ionized disk predicts 
Fe K$\alpha$ line energies of $\ge6.6$ keV, for which 
there is little evidence (e.g., Nandra et al. 1997). 
On the other hand, there may be a contribution to 
the high-energy side of the Fe K$\alpha$ line from warm 
scattering material located above the torus, which would reduce  
the EQW of the disk line even further.

X-ray satellites like $AXAF$ and $Astro-E$, with much better resolving 
power, will be able to deconvolve the line and thus  
will provide the ultimate test of whether 
the Fe K$\alpha$ lines of intermediate type Seyfert galaxies 
are truly complex with distinct, unresolved cores.  

\section{Conclusions}

We demonstrate that the Fe K$\alpha$ emission lines for four 
X-ray selected Seyfert type 1.9 and 2 galaxies  
are well described with a model that consists of a superposition 
of emission from an accretion disk viewed at an intermediate angle of 
$\sim 48^{\circ}$ and a distinct, 
unresolved emission feature with an equivalent width of $\sim 60$\,eV that
can arise in the obscuring torus.  This model is entirely consistent with
current unified schemes of Seyfert galaxies and therefore does not 
require severe misalignments between the accretion disk and the torus.

\vskip0.6in

We thank Julian Krolik for reading this 
manuscript and providing helpful comments and insight. 
KAW acknowledges the support of NASA LTSA grant 
NAG5-3504.  CSR acknowledges support from NSF grant 
AST9529175, and NASA LTSA grant NAG5-6337.

\clearpage

\begin{deluxetable}{lcccccccc}
\tablenum{1}
\small              
\tablewidth{0pt}
\tablecaption{ASCA Results for Power-law and Narrow Gaussian Fits$^1$
\label{tab:table1}}
\tablecolumns{9}
\tablehead{\colhead{Galaxy} & \colhead{z} & 
 \colhead{$\chi^2_{\rm PL}/{\nu}^2$} &
 \colhead{$N_{\rm H}^3$} & \colhead{$\Gamma^4$} &
 \colhead{EQW$_{\rm N}^5$} & \colhead{$\chi^2_{\rm G}/{\nu}^6$} & 
 \colhead{$\Delta\chi^2$\,$^7$} & \colhead{F$^8$} 
}
\startdata
NGC 526A & 0.0192 & 1509/1512 & 12.7$^{+0.9}_{-0.8}$ & 1.65$^{+0.02}_{-0.03}$ 
      & 113$^{+9}_{-25}$ & 1453/1511 & 56 & 58.3 \nl
NGC 2110 & 0.0076 & 815/757 & $33.9\pm1.4$ & 1.46$^{+0.03}_{-0.04}$ 
      & 140$^{+18}_{-17}$ & 748/756 & 67 & 67.7 \nl
MCG$-$5-23 & 0.0083 & 1435/1247 & 14.1$^{+0.6}_{-0.7}$ & 1.73$\pm0.02$ 
      & 93$\pm10$ & 1349/1246 & 86 & 79.6 \nl
NGC 7314 & 0.0047 & 1581/1461 & 5.7$^{+0.6}_{-0.8}$ & 1.85$\pm0.03$ 
      & 120$^{+12}_{-22}$ & 1530/1460 & 51 & 48.6 \nl
\tablenotetext{1}{The data and selection criteria are identical to
those used by T98.  Data are fitted from 2 to 10 keV and errors
are 1-$\sigma$.  Spectra are grouped to have $>$20 counts per bin.
The Fe K$\alpha$ line energy is fixed at 6.4 keV in 
the galaxy rest frame and the line width is fixed at 0.01 keV.}
\tablenotetext{2}{Value of the $\chi^2$ statistic divided by the
number of degrees of freedom for the power-law (PL) fit (no Gaussian).}
\tablenotetext{3}{Absorbing column density for the PL plus 
narrow Gaussian fit in units of 10$^{21}$ cm$^{-2}$.}
\tablenotetext{4}{Photon index for the PL plus Gaussian fit.}
\tablenotetext{5}{Equivalent width of the Gaussian in units of eV.}
\tablenotetext{6}{Value of the $\chi^2$ statistic divided by the
number of degrees of freedom for the PL plus Gaussian fit.}
\tablenotetext{7}{Change in $\chi^2$ for PL plus Gaussian fit compared 
to the PL fit (col.\ 3).}
\tablenotetext{8}{F-statistic (Bevington 1969).}
\enddata
\end{deluxetable}

\clearpage
         
\begin{deluxetable}{lcccccccc}
\tablenum{2}
\footnotesize       
\tablewidth{0pt}
\tablecaption{Results for Disk-line Fits$^1$   
\label{tab:table2}}
\tablecolumns{9}
\tablehead{\colhead{Galaxy} & \colhead{$N_{\rm H}$} &
  \colhead{$\Gamma$} &  
  \colhead{$\theta^2$} & \colhead{EQW$_{\rm D}^3$} & \colhead{$\chi^2/\nu^4$} &
  \colhead{$\Delta\chi^2$\,$^5$} & \colhead{F$^6$} & \colhead{Flux$^7$}
}
\startdata
NGC 526A & $13.2\pm0.9$ & $1.69\pm0.03$ &
  34$^{+4}_{-3}$ & 230$^{+30}_{-40}$
  & 1454/1510 & -1 & \nix  & 3.6 \nl
NGC 2110 & 34.3$^{+1.4}_{-1.6}$ & $1.51\pm0.04$ 
    & 33$^{+3}_{-4}$ & 324$^{+46}_{-40}$ & 744/755 & 4 & 4.1 & 3.0 \nl
MCG$-$5-23 & 14.9$^{+0.7}_{-0.6}$ & 1.79$^{+0.02}_{-0.03}$ 
    & 38$^{+5}_{-4}$ & $270\pm30$ & 1326/1245 & 23 & 21.5 & 8.7 \nl
NGC 7314 & 7.3$^{+0.8}_{-0.9}$ & 1.95$^{+0.03}_{-0.04}$ 
   & 45$^{+2}_{-3}$ & $360\pm50$ & 1512/1459 & 18 & 17.3 & 3.6 \nl
\tablenotetext{1}{Disk emissivity index $q$ is fixed at $-2.5$.
Errors are 1-$\sigma$. The outer and inner disk radii are fixed at 
1000$r_g$ and 6$r_g$ ($r_{g} = GM/c^2$).  Line energies are 6.4 keV in the galaxy
rest frame. The units for $N_{\rm H}$ are 10$^{21}$ cm$^{-2}$.}
\tablenotetext{2}{Inclination of the disk normal to our 
line of sight in units of degrees ($\theta=0^{\circ}$ is face-on).}
\tablenotetext{3}{The equivalent width of the disk line in units of eV.}
\tablenotetext{4}{Value of the $\chi^2$ statistic divided by the
number of degrees of freedom.}
\tablenotetext{5}{Change in $\chi^2$ for this fit compared to the narrow 
Gaussian fit (Table 1).}
\tablenotetext{6}{F-statistic.}
\tablenotetext{7}{The 2 to 10 keV observed flux in units of 
10$^{-11}$ ergs cm$^{-2}$ s$^{-1}$.}
\enddata
\end{deluxetable}

\clearpage

\begin{deluxetable}{lcccccccc}
\tablenum{3}
\small              
\tablewidth{0pt}
\tablecaption{Results for Disk-Line Plus Narrow Gaussian Fits$^1$
\label{tab:table3}}
\tablecolumns{9}
\tablehead{\colhead{Galaxy} &
 \colhead{$N_{\rm H}$} & \colhead{$\Gamma$} & 
 \colhead{$\theta^2$} & \colhead{EQW$_{\rm D}^3$} & 
 \colhead{EQW$_{\rm N}^3$} & \colhead{$\chi^2/\nu^4$} &
 \colhead{$\Delta\chi^2$\,$^5$} & \colhead{F$^6$}
}
\startdata
NGC 526A & $13.3\pm0.9$ & 1.68$^{+0.03}_{-0.02}$  
    & 42$^{+13}_{-6}$ & $120\pm40$   
    & 64$^{+17}_{-19}$ & 1444/1509 & 10 & 10.4 \nl
NGC 2110 & 35.3$^{+1.4}_{-1.6}$ & 1.54$^{+0.05}_{-0.04}$ 
    & 50$^{+3}_{-4}$ & 246$^{+57}_{-28}$ & 81$^{+16}_{-21}$ 
    & 728/754 & 16 & 16.5 \nl 
MCG$-$5-23 & 15.5$^{+0.6}_{-0.7}$ & 1.81$^{+0.02}_{-0.03}$ 
    & $52\pm4$ & 235$\pm35$ & 50$^{+9}_{-11}$ 
    & 1309/1244 & 17 & 16.2 \nl
NGC 7314 & 7.2$^{+0.7}_{-1.0}$ & 1.94$^{+0.02}_{-0.04}$ & 46$^{+5}_{-2}$ 
    & $250\pm50$ & 46$^{+19}_{-15}$ & 1505/1458 & 7 & 6.8 \nl
\tablenotetext{1}{Errors are 1-$\sigma$. The disk emissivity 
index $q$ is fixed at $-2.5$; $r_i = 6r_g$, $r_o = 1000r_g$.
Line energies are 6.4 keV in the galaxy rest frame.}
\tablenotetext{2}{Disk inclination in units of degrees.}
\tablenotetext{3}{Equivalent widths of the disk line and 
narrow Gaussian ($\sigma_{\rm N} = 0.01$ keV).}
\tablenotetext{4}{Value of the $\chi^2$ statistic divided by the
number of degrees of freedom.}
\tablenotetext{5}{Change in $\chi^2$ compared to the disk-line
fits (Table 2).}
\tablenotetext{6}{F-statistic.}
\enddata
\end{deluxetable}

\clearpage

\begin{deluxetable}{lccccc}
\tablenum{4}
\small              
\tablewidth{0pt}
\tablecaption{Comparing Our Results with T98$^{1,2}$ 
\label{tab:table4}}
\tablecolumns{6}
\tablehead{\colhead{} & \multicolumn{2}{c}{This paper} &
 \multicolumn{2}{c}{T98} & \colhead{} \nl 
\colhead{Galaxy} & \colhead{$\theta$} & \colhead{EQW$_{\rm D}$} &
  \colhead{$\theta$} & \colhead{EQW$_{\rm D}$} &
  \colhead{$b/a^3$} \nl
\colhead{} & \colhead{($^{\circ}$)} & \colhead{(eV)} & 
\colhead{($^{\circ}$)} & \colhead{(eV)} & \colhead{}
}
\startdata
NGC 526A &  42$^{+13}_{-6}$ & $120\pm40$ 
 & 13$^{+6}_{-13}$ & $281\pm56$ & 0.7 \nl
NGC 2110 &  50$^{+3}_{-4}$ & 246$^{+57}_{-28}$  
 & 15$^{+9}_{-7}$ & 289$^{+59}_{-52}$ & 0.79 \nl
MCG$-$5-23 & 52$\pm$4 & 235$\pm$35 
 & 33$^{+11}_{-4}$ & 362$^{+94}_{-43}$ & 0.45 \nl
NGC 7314 &  46$^{+5}_{-2}$ & $250\pm50$ 
 & 17$^{+6}_{-7}$ & $417\pm82$ & 0.43  \nl
$\mu$+$\sigma$ & $48\pm4$ & $213\pm62$ & $20\pm9$ 
 & $337\pm64$& \nix \nl
\tablenotetext{1}{Line energies are fixed at 
the rest energy of 6.4 keV; the disk emissivity index is 
$q=-2.5$; the inner and outer disk radii 
are $6r_g$ and $1000r_g$, respectively ($r_{g} = GM/c^2$).} 
\tablenotetext{2}{The equivalent width for the narrow 
Gaussian component is a free parameter for our fits and constrained 
to $<50$ eV for the T98 fits.  We derive 
$<{\rm EQW}_{\rm N}> = 60$ eV.}
\tablenotetext{3}{Axial ratio of host galaxy.}
\enddata
\end{deluxetable}

\clearpage
\vfill\eject

\clearpage 

\begin{figure}
\epsscale{0.80}
\plotfiddle{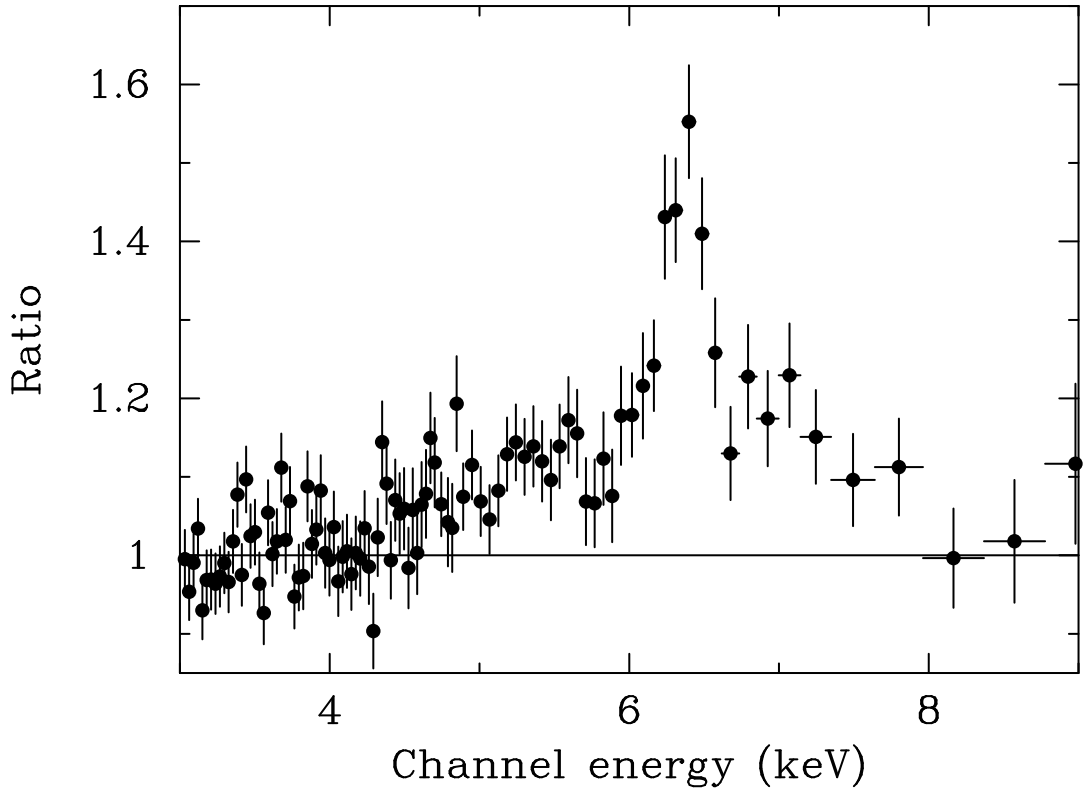}{150pt}{-90}{70}{70}{-285}{380}
\caption[ ]{Average ratio of the $ASCA$ Solid-State Imaging 
Spectrometer (SIS) data to the best-fitting
continuum models.  Individual spectra were corrected for
redshift and the data above and below the Fe K$\alpha$ line
(2 to 4.5 keV and 8 to 10 keV) were fitted with a power law 
model to estimate the continuum level.  
The ratios of the data to this model for each galaxy were then 
calculated and averaged together with {\it xspec}
(see also T98 Fig. 2, group B).
\label{fig:data}
}
\end{figure}

\clearpage

\begin{figure}
\epsscale{0.75}
\plotone{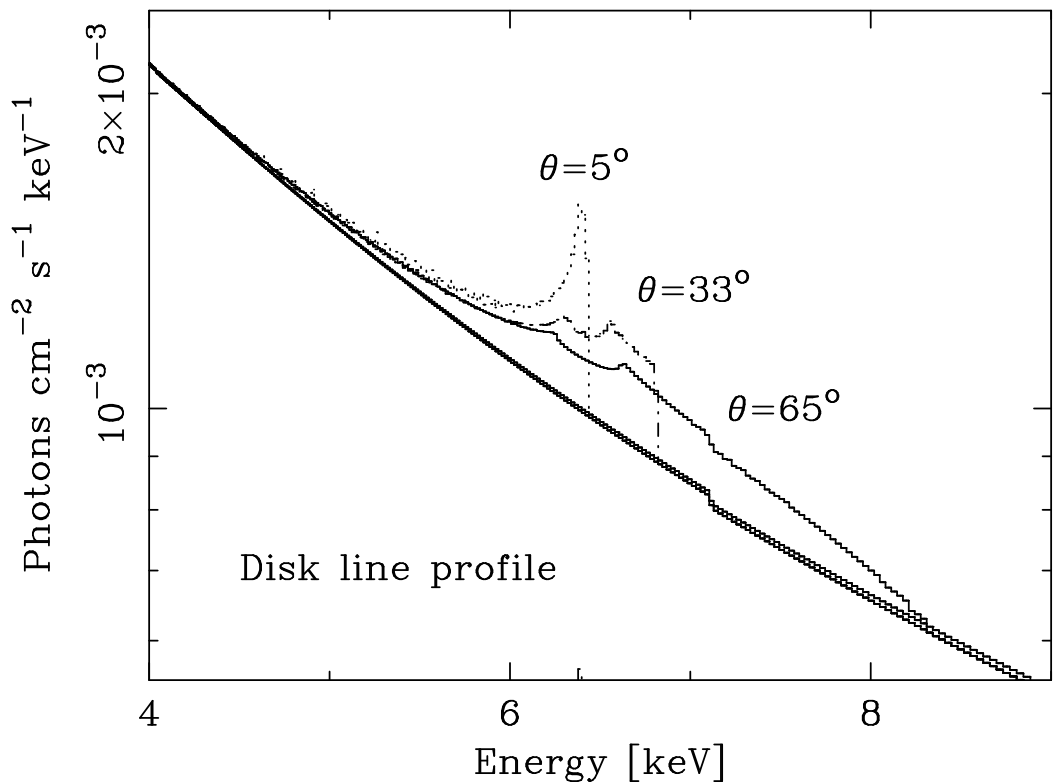}
\caption[ ]{Theoretical line profiles from a
relativistic accretion disk (Fabian et al.\ 1989) to illustrate
how the profile changes as the inclination of 
the disk normal to our line of sight increases.  Inclinations 
are 5$^{\circ}$ (dotted), 33$^{\circ}$ (dashed), and 65$^{\circ}$ (solid). 
Other line parameters are E$_{\rm peak}$=6.4 keV,
disk emissivity index, $q = -2.5$ (r$^q$), inner radius, $r_{i}=6r_g$,
and outer radius, $r_{o}=1000r_g$ ($r_g=GM/c^2$).
\label{fig:lines1}
}
\end{figure}

\clearpage

\begin{figure}
\epsscale{0.80}
\plotfiddle{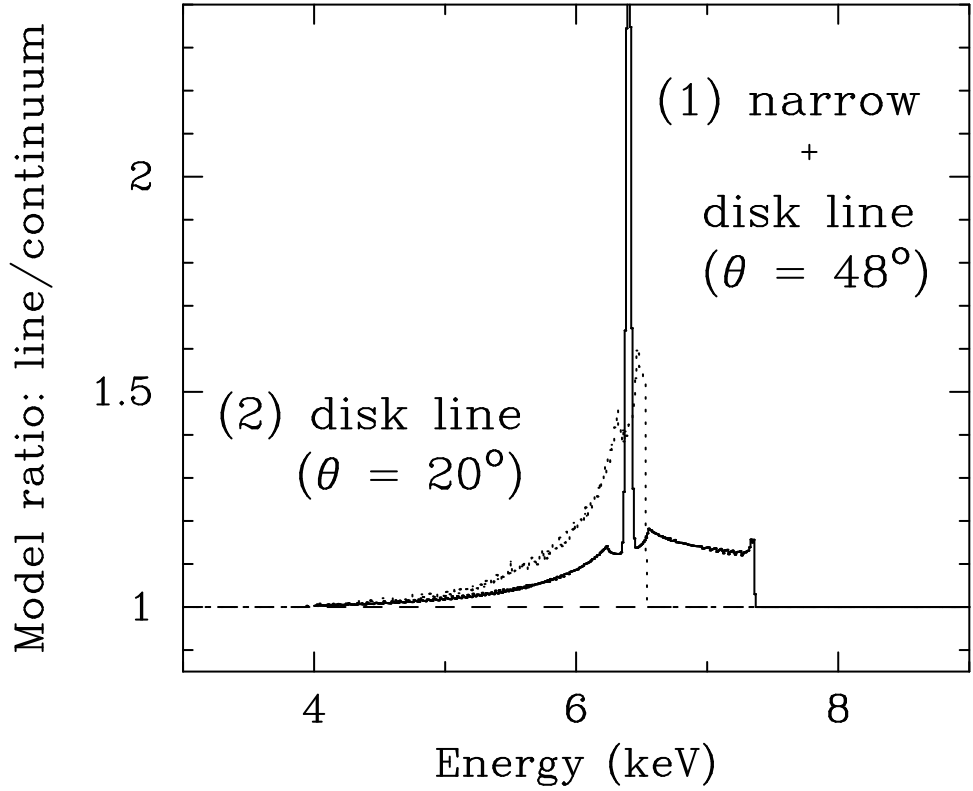}{150pt}{-90}{70}{70}{-290}{380}
\caption[ ]{Theoretical line profiles for 
(1) an unresolved line
superimposed on a line from an accretion disk viewed at 
48$^{\circ}$ (solid line) and (2) a line
from an accretion disk viewed at an inclination 
of 20$^{\circ}$ (T98, dotted line). 
This figure was produced by calculating the 
model spectrum (line plus power-law continuum) and dividing 
by the continuum model. 
\label{fig:lines2}
}
\end{figure}

\end{document}